\magnification=1200 \baselineskip=13pt \hsize=16.5 true cm \vsize=20 true cm
\def\parG{\vskip 10pt} \font\bbold=cmbx10 scaled\magstep2

\centerline{\bbold Evolutionary Computer Simulations}\parG
\centerline{Paulo Murilo Castro de Oliveira}\parG
Instituto de F\'\i sica, Universidade Federal Fluminense\par
av. Litor\^anea s/n, Boa Viagem, Niter\'oi RJ, Brasil 24210-340\par
PMCO @ IF.UFF.BR

\vskip 0.6cm\leftskip=1cm\rightskip=1cm 

{\bf Abstract}\parG

	Computer modelling for evolutionary systems consists in: 1) to
store in the memory the individual features of {\bf each} member of a
large population; and 2) to update the whole system repeatedly, as time
goes by, according to some prescribed rules (reproduction, death, ageing,
etc) where some degree of randomness is included through pseudo-random
number sequences. Compared to direct observation of Nature, this approach
presents two distinguishing features. First, one can {\bf follow} the
characteristics of the system in real time, instead of only observing the
current, static situation which is a long-term consequence of a remote
past completely unknown except for some available fossil snapshots. In
particular, one can {\bf repeat} the whole dynamical process, starting
from {\bf the same} initial population, using {\bf the same} randomness,
changing only some minor contingency during the process, in order to study
its long-term consequences. Second, evolution necessarily follows a {\bf
critical} dynamics with {\bf long-term memory} characteristics, equivalent
to the long-range correlations responsible for the well known universality
properties of static critical phenomena. Accordingly, some strong
simplifications can be applied, allowing one to obtain many
characteristics of real populations from toy models easily implementable
on the computer.

\leftskip=0pt\rightskip=0pt\vfill

\eject

	What nowadays we call {\bf evolution} --- the theory describing
how living beings behave under permanent changes, the whole dynamical
system {\bf never} reaching any final, absorbing state --- was introduced
by Jean Lamarck [1] two centuries ago. He also invented the name
``biologie''. According to him, the many currently observed biological
species are not static, independent entities, but derive from ancient ones
after small modifications accumulated during {\bf very long} times.
Lamarck fell into disgrace under the religious/conservative power
dominating his lifetime. Reference [2] presents an amusing and intriguing
discussion concerning these subjects. The mechanism leading to this
forever-changing scenario is Charles Darwin's natural selection [3]. Each
individual inherits the characteristics of its parents, with some small
modifications, being similar but not identical to neither the parents nor
the siblings. Among the latters, the ones better adapted to survive under
the current environment are more likely to breed, their offspring carrying
the naturally selected traits from their grand-parents.

	A simple way to interpret evolution is to imagine the
high-dimensional space of all possible living forms. Each individual is a
point in this space, and their offspring are neighbouring points.
Considering asexual reproduction, for simplicity, each offspring can also
produce its own offspring, an so on, according to a branching dynamical
process. Some selected branches of the resulting tree continue to grow
further, whereas others are dangling ends without offspring, avoiding
population explosion. Under a large-scale overview, a snapshot of this
space of all possible living forms shows a very sparsely, highly
inhomogeneous occupation pattern: many neighbouring occupied points form a
cloud localised in some small region, separated from other equivalent
clouds. Each such a cloud is a species. Following such a cloud during a
short-term time interval, i.e. some few generations, its many occupied
points are continuously replaced by other neighbouring points: the whole
cloud seems to stay immobile. Only under a long-term point of view, i.e.
after a large number of generations, one can observe the extremely slow
movement of the cloud as a whole. Tracing back the movement of two
nowadays separated clouds, one would discover that both derive from the
same ancient, possibly no-longer existing, parent cloud, the so-called
speciation process. Thus, evolution occurs according to a branching
process not only under the local point of view of each individual and its
offspring, but also under the much larger scale of species as a whole.

	Under a coarse-grained point of view, the same space of all
possible living forms can be re-interpreted: now each cloud (species)
corresponds to a single slowly moving Point (capital P to distinguish it
from a single individual). Neighbouring occupied Points form a Cloud, i.e.
a set of neighbouring species, called a genus. Stretching once more the
scale, each genus can be interpreted as a single PPoint (double PP, since
my keyboard does not provide anything larger than capital letters) within
the same space of all possible living forms. A CCloud of PPoints is called
a family. An order is a CCCloud of PPPoints. A class is a CCCCloud of
PPPPoints. Note that each step along this hierarchical sequence of
scalings involves not only larger and larger scales of length (distances
measured within the space of all possible living forms), but also larger
and larger scales of time!

	For statistical physicists, the above description has certainly a
feeling of ``d\'ej\`a-vu''. It is just the basic reasoning used in order
to study critical, scale-free phenomena, the fundamental concept behing
the renormalisation-group theory [4] for which Kenneth Wilson was awarded
with the 1982 Nobel prize. These phenomena are normally described by
power-law decays of the various correlations as functions of {\bf both}
length and time. Indeed, there is a lot of evidence in favour of this
scenario also governing biological evolution (see, for instance, [5] and
original references therein). However, the above iterative-scaling
reasoning (individual -- species -- genus -- family -- order -- class) has
only an illustrative purpose.

	An argument showing that evolution necessarily follows a {\bf
critical dynamics} was presented in [6]. In short, it is as follows (six
paragraphs). One always needs to separate some part of the universe (a
species, a geographic region, etc), in order to study its evolutionary
behaviour. This artificially separated part cannot be considered a closed
object, because it is always under some influence of the rest of the
universe, its environment. Thus, the system under study must necessarily
be considered as an {\bf open dynamic system}. Some ingredients (mass,
energy, information, etc) feed continuously the system, and in some way
are processed therein. Also, as a by-product of this procedure, the same
kind of ingredients are continuously thrown out. That is why the dynamic
evolution of such a system is always dissipative, irreversible: the
would-be final situation, i.e. the attractor, is a low-dimensional object,
a null measure set compared to the whole initially available set of
possibilities. Among the many directions of the high-dimensional space of
all possible living forms, only a few remain available after the system is
trapped forever inside its tiny attractor. After that, all other
directions are {\bf extinct}. This is not a profitable feature, in what
concerns evolution by natural selection, which demands the {\bf eternal}
possibility of visiting new forms.

        The difference between the so-called regular or chaotic behaviours
is the following. Starting from two slightly distinct initial conditions,
the distance $\Delta$ from each other will evolve as

$$\Delta(t) \sim {\rm e}^{\lambda t}$$

\noindent as the time $t$ goes by. The so-called Lyapunov exponent
$\lambda$ may be negative, which corresponds to the regular case: both
trajectories eventually converge to a single one, a simple attractor.
Otherwise, for positive values of $\lambda$, the system is chaotic: both
trajectories diverge from each other inside a complicated object (anyway a
low-dimensional one), usually called a strange attractor. In both cases,
regular or chaotic, after a {\bf finite} transient time $\tau \sim \vert
1/\lambda\vert$ the system becomes trapped, loosing forever its ability to
explore the whole set of possibilities. Again, evolution is not compatible
with this scenario.

        The question now is: how to preserve diversity of options, i.e.
the ability to reach any part of the high-dimensional space of
possibilities, while the dynamic process itself traps the system more and
more close to a low-dimensional destiny? It is not a simple matter of
loosing some fraction of the whole space. It is worse than that: one
looses entire directions! Nature is smart enough to avoid this catastrophe
by postponing it {\bf forever}: Nature ``chooses'' $\lambda = 0$, the
complex, critical case. The above equation misses sub-dominant terms
(not-shown), which in this case become the dominant ones imposing a much
slower decaying rate. Normally they are power-laws, sometimes
slower-yet-decaying relations, anyway lacking any typical time (and
length) scale. The difference is not a mere quantitative one, it is rather
qualitative: the difference between any finite time interval, no matter
how long it could be, and {\bf eternity}. In order to better understand
this concept, consider an analogy with a radioactive decay described just
by the above equation: $\tau \sim \vert 1/\lambda\vert$ is the so-called
{\bf waiting time}, the time one needs to wait for the decay of some
particular nucleus, in average. After some finite time interval (say
$2\tau$, $4\tau$) the whole sample looses its radioactivity. In other
cases for which this characteristic scale is lost, i.e. $\lambda = 0$, the
system's activity {\bf never} ceases. This is just what evolution needs.

	Here, the precise mathematical meaning of the word {\bf never} is
the following. For an exponential decay, the waiting time does not depend
on the system's size: one can make the system larger and larger, and the
waiting time will stay at the same {\bf finite} value. For the scale-free
case $\lambda = 0$, however, the waiting time grows indefinitely for
larger and larger systems. In other words, the lifetime of a critical
system is limited only by its size. The lack of any typical size scale
implies the corresponding lack of any typical time scale, and vice-versa.
For purists, another reasoning leading to the same conclusion follows. In
order to keep finite the minority number of individuals carrying some rare
trait, avoiding its extinction, one needs to keep alive a large enough
whole population. Tolerating some fixed, large-but-finite extinction time
$T$, one needs to keep a population which increases exponentially with the
extinction rate $\vert \lambda \vert$. For the scale-free case $\lambda =
0$, however, it increases according to some other slower-than-exponential
rate, say a power-law, which requires a not-so-large population. In this
case, extintion can always be postponed beyond $T$ only by keeping some
moderate population, i.e. effectively {\bf forever}. A nice example of how
Nature adopts this strategy is the extremely slow rate of extinction
observed for recessive diseases [7,6], where diploidism plays a major
role.

        Nature does not choose anything. In reality, natural selection
itself tunes to the critical, complex case $\lambda = 0$: only those
situations avoid their own extinction, among all other possibilities which
die out along the process. That is why so many examples of fossil data
seem to agree with this scenario: the other possibilities, if any, did not
survive long enough. Also, the Lyapunov exponent $\lambda$ is not just
one, their number coincides with the system's dimensionality. Natural
selection tunes as many of them as possible to the critical value $\lambda
= 0$.

	Another interesting interpretation of the same concept, namely the
preservation of diversity, is as follows. In separating the system under
study from its environment, only an artificial working procedure, one
needs to realise that the environment itself is not constant, that it also
evolves (normally under a slower rate). The attractor of the dynamical
process can be viewed as the {\bf current best} option. However, future
environment modifications will slightly displace this optimum to
neighbouring positions. Thus, evolution needs to keep the current
attractor's neighbourhood always populated, in order to fit these
unpredictable future displacements. In other words, some degree of
diversity {\bf near but outside} the current attractor must be kept
forever. This feature is automatically provided by critical dynamics,
according to which the system continuously approaches its attractor but
never reaches it. Evolution is not an eugenic optimisation process where
only the current ``best'' is selected to survive: this would forbid
natural selection itself to proceed, due to the consequent lack of future
options. Any non-critical dynamics, where the ``best'' form is reached
after some finite transient time, rules out evolution. Indeed, according
to [8], eugenics leads to extinction.

        Probably, the reader is not confortable with the expression
``space of all possible living forms'', in which all previous reasoning is
based. How to define such an object? The first to address this question
was Gregor Mendel [9], the founder of genetics, in his famous two lectures
of 1865. By breeding sexually reproducing plants, he discovered that
particular individual traits are inherited as a whole, all or nothing, yes
or not, 1 or 0. Each individual inherits half of its traits from the
father, half from the mother. Mendel provided the basic ``software''
theory for genetics. His work was completely forgotten for decades. Much
later, chromosomes and diploidism were discovered, confirming his original
findings and providing the ``hardware'' for parent's inheritance
mechanism. Mendel's ``traits'' were then physically associated with genes,
small pieces of the chromossomes. Since then, it becomes clear that
characteristics acquired during the lifetime of an individual are not {\bf
genetically} passed on to its offspring. Before that, missing this
molecular-based genetic mechanism, both Lamarck and Darwin did not ruled
out the possibility of passing acquired traits to the offspring (which
indeed occurs, but not genetically, except for rare cases of damaged germ
cells).

        Thanks to Mendel's idea, one can easily conceive an array of bits
1 or 0, each one meaning whether some particular gene is present or not in
a particular individual belonging to the population under study. Thus,
each conceivable individual is represented by such an array, its total
number of bits being the dimension of the quoted ``space of all living
forms''. Indeed, sexually reproducing individuals must be represented by
two bit-arrays each, one inherited from the father, other from the mother.
Also, the same gene can appear in a few different forms, alleles: in order
to deal with this further complication, one can simply allocate more than
one bit per gene. Anyway, our ``space of all living forms'' is now well
defined: its elements are bit-strings, each one corresponding to a
possible existing individual (its ``genome''). The population dynamics
corresponds to the evolution of a variable number of these individuals
which are alive at each time step $t$. The dynamic evolution follows some
rules concerning birth, death, interaction with the environment, etc. {\bf
All} biological issues are included in such rules. The first mathematical
description of evolution based on this bit-string representation was the
famous Eigen model [10].

	At this point, after a long and dangerous digression over other
fields of knowledge, concerning which I hope the reader did not notice my
complete ignorance, I am back to the main subject of the title. Being a
computer physicist without access to powerful computers, I learned many
tricks useful to save computer time and memory. One particularly important
strategy is to operate directly on the bits of each computer word (a
sequence with 32 bits), through bitwise, parallel operations [11]. These
tricks allow one to perform incredible, otherwise impossible jobs.
Computer modelling consists simply in programming such rules on the
computer, let them work as time goes by, and observe the outcome. As
biological issues are an enormous source of complications, one needs to
simplify these rules. In reality, one plays with toy models, opening the
way for criticisms related to reductionism. On the other hand, they are
very convenient in order to analyse which ingredients are really important
and which are not, concerning particular population features. Within a
complex subject as biology, this approach could contribute with some light
for questions particularly difficult to answer by other means, moreover
due to the advantages of following the evolution in real time and the
possibility of repeating the whole history under controlled contingencies.

	An example is a conjecture raised by Stephen J. Gould [12]: would
some contingency have occurred a long time ago, even a small one, slightly
different than the historical truth, then we could observe nowadays a
different scenario concerning the current existing species. The true
evolutionary tree historically followed since the onset of life on Earth
is not the only possible one, many others could also have occurred.
Obviously, this issue cannot be verified by observation, however, it can
easily be simulated on computers in the way described above. There is a
set of arguments against evolution based on probabilistic estimatives of
the time needed to reach the enormous biological complexity we can observe
nowadays, resulting in values many orders of magnitude larger than the
Universe age! However, one needs at least to divide these estimatives by
the number of possible alternative scenarios, not just by the single one
we observe today. How many they are? This is a problem very difficult or
impossible to be solved by other means than computer simulations. Another
way to understand this puzzle is to consider only a small step of the
evolutionary time, a snapshot of the evolutionary tree and all its
subsequent potentially possible scenarios a little bit later: among them,
many can be discarded immediately, presenting no growing branches at all.
Thus, only the remainders which present some possibility of further
growing need to be taken into account, not the whole potential set. This
dissipative feature goes on at each new time step, cumulatively. It
challenges the estimatives of the age of life on Earth obtained by a
direct comparison between the currently observed scenario and some
plausible starting condition, disregarding all dissipated possibilities in
between.

	In order to construct an evolutionary computer model, the first
ingredient to be taken into account is the reproductive behaviour, for
instance: the bit-string of the parent is copied into its offspring, with
some possible random mutations (to flip a small number of bits). For a
sexual population, with two parallel bit-strings per individual, first one
cuts both bit-strings of the mother at the same random position, and then
form one bit-string by joining two complementary pieces to be inherited by
the offspring. The other is obtained by performing the same on the
father's genome. The second ingredient concerns death, for instance: the
death roulette randomly kills individuals at every new time step, in order
to keep constant the population. Further ingredients can be included. For
instance, the mutation rate can vary for different genes (different
positions along the bit-string). One can also give some preference for the
position where the crossing is performed. Also, diverse correlations
between different genes can be included. Birth and death rates can be
genome-dependent. Geographic concerns and migrations can also be adopted.
I am a very lucky guy, because I know how to efficiently implement such
rules on the computer, and also because one has a virtually infinite
number of new biological ingredients to be tested. Nowadays, for every
physics problem there are a thousand physicists working behind, while for
every thousand problems in biology there is one biologist [13]. I invite
everybody to contribute.

	I will finish with a single example of an evolutionary computer
model, which was invented in our Institute [14]. For a review, see [15]
and [16]. A further fundamental ingredient is ageing. One can keep the age
of each alive individual, updating it every new time step, i.e. every new
``year''. Instead of the whole genome, one can store only a reduced
chronological projection of it, say a 128-bit string: each bit corresponds
to one year, during the individual's lifetime. We will call this
bit-string ``genome'', anyway, since it is not modified during life. A bit
0 at position $a$ means that no disease will appear at this age.
Otherwise, a harmful genetic disease corresponds to a bit 1: for instance,
an individual starting to be affected by Alzheimer disease since 60 years
old has a bit 1 at position $a = 60$. Note that Alzheimer disease starts
to be harmful only after 60, but it was already present in the
individual's genome since birth. A child suffering from Down's syndrome
has a bit 1, say, at position $a = 2$. At each new year, every alive
individual whose age is above some minimum reproduction age $R$, say $R =
15$, is allowed to breed $B$ offspring, say $B = 1$, without or with sex
[17]. Also each individual with more than $T$ genetic diseases accumulated
up to its current age, say $T = 3$, dies. Another set of individuals also
dies at random, independently of age or genome, in order to mimic
non-genetic deaths avoiding population explosion.

	This model makes concrete the evolutionary theory of ageing [18].
According to this theory, we age due to inheritance from our ancestors:
would they have too much genetic diseases programmed in their genomes to
occur at youth, then they would die before breeding. Thus, our true
ancestors had their genomes ``clean'' enough (without or with only few
genetic diseases) {\bf up to} their reproduction age. However, they could
be our ancestors even presenting very ``dirty'' genomes {\bf after} their
reproduction age. Being descendents of them, so we are: almost free of
genetic diseases before the reproduction age. That is why we start to
suffer from more and more diseases when we become old, i.e. after the
reproduction age. A guy like Dietrich Stauffer, who has no known
offspring, is supposed to live forever, according to his own (wrong)
interpretation of the theory. A striking consequence of this theory is: a
species which, besides the minimum reproduction age, has also a maximum
reproduction age $M$ (for menopause), evolves to a situation such that
nobody survives beyond age $M$. This is just the case of the pacific
salmon, for instance. This behaviour is nicely reproduced [19] by the
Penna model, simply by introducing this new ingredient: a maximum
reproduction age. Women present menopause around 50 years old, contrary to
men who are in principle able to reproduce until death. Thanks to men,
women were allowed by Nature to live beyond 50. Is menopause explainable
from evolutionary arguments? According to Williams [20] the answer is yes.
His theoretical argument concerns a compromise between child care and
reproduction risk for the mother: it would be better for the mother to
stop reproduction at some age, avoiding the risk of death during future
childbirths, in order to be able to take care of the already born
offspring. Indeed, including these two ingredients, namely child care and
reproduction risk, a menopause age naturally arises from the Penna model
[21]. According to data collected by Sir Thomas Perls and colleagues [22],
longevity is inheritable. Again, the Penna model reproduces quite well
these data [23].

	Many similar models consider characteristics other than ageing,
some of them in [24].

\vskip 25pt
\centerline{\bf Aknowledgements}\parG

	I am indebted to my many collaborators concerning this subject,
specially S. Moss de Oliveira, T.J.P. Penna, D. Stauffer, A.T. Bernardes
and J.S. S\'a Martins. I am also indebted to Jean Lobry and Leny
Cavalcante for references [2] and [9], respectively. Financial support
came from Brazilian agencies CAPES, CNPq and FAPERJ.

\vskip 40pt
\centerline{\bf References}\parG

\item{[1]} J. Lamarck, {\sl Hydrog\'eologie} (1802); {\sl Recherche sur
l'organisation des corps vivants} (1802); {\sl La Phylosophie Zoologique}
(1809).\par

\item{[2]} A. Langaney, {\sl La Phylosophie $\dots$ Biologique}, Belin,
Paris (1999).\par

\item{[3]} C. Darwin, {\sl On the Origin of Species by Means of Natural
Selection}, Murray, London (1859).\par

\item{[4]} Although we still lack a complete theoretical description
involving the time, its successful static version is already
available:\par

\item{} K.G. Wilson and J. Kogut, {\it Phys. Rep.} {\bf 12C}, 75 (1974);
K.G. Wilson, {\it Sci. Am.} {\bf 241}, 140 (August 1979).\par

\item{[5]} P. Bak, {\sl How Nature Works: the Science of Self-Organized
Criticality}, Oxford University Press (1997); S.A. Kauffman, {\sl At Home
in the Universe}, Oxford University Press (1995); {\sl Origins of Order:
Self-Organization and Selection in Evolution}, Oxford University Press
(1993).\par

\item{[6]} P.M.C de Oliveira, {\it Theory in Biosciences} {\bf 120}, 1
(2001), also in COND-MAT 0101170.\par

\item{[7]} A. Jacquard, {\sl \'Eloge de la Diff\'erence: la G\'en\'etique
et les Hommes}, \'Editions du Seuil, Paris (1978).\par

\item{[8]} S. Cebrat and A. Pekalski, {\it Eur. Phys. J.} {\bf B11}, 687
(1999).\par

\item{[9]} G. Mendel, {\it Verhandlungen des Naturforschenden Vereins}
(1866), also in http:\par
\item{} //www.netspace.org/MendelWeb/MWNotes.html.\par

\item{[10]} M. Eigen, {\it Naturwissenschaften} {\bf 58}, 465 (1971); M.
Eigen, J. McCaskill and P. Schuster, {\it Adv. Chem. Phys.} {\bf 75}, 149
(1989).\par

\item{[11]} P.M.C de Oliveira, {\sl Computing Boolean Statistical Models},
World Scientific, Singapore London New York, ISBN 981-02-0238-5
(1991).\par

\item{[12]} S.J. Gould, {\sl The Wonderful Life}, Norton, New York
(1989).\par

\item{[13]} R. Koberle, private communication.

\item{[14]} T.J.P. Penna, {\it J. Stat. Phys.} {\bf 78}, 1629 (1995).\par

\item{[15]} S. Moss de Oliveira, P.M.C. de Oliveira and D. Stauffer, {\sl
Evolution, Money, War and Computers: Non-Traditional Applications of
Computational Statistical Physics}, Teubner, Stuttgart Leipzig, ISBN
3-519-00279-5 (1999).\par

\item{[16]} D. Stauffer, P.M.C. de Oliveira, S. Moss de Oliveira, T.J.P.
Penna and J.S. S\'a Martins, {\it Anais Acad. Bras. Cienc.} {\bf 73}, 15
(2001), also in COND-MAT 0011524; S. Moss de Oliveira, D. Alves and J.S.
S\'a Martins, {\it Physica} {\bf A285}, 77 (2000); S. Moss de Oliveira,
{\it Physica} {\bf A257}, 465 (1998).\par

\item{[17]} S. Moss de Oliveira, P.M.C. de Oliveira and D. Stauffer, {\it
Braz. J. Phys.} {\bf 26}, 626 (1996); D. Stauffer, P.M.C. de Oliveira, S.
Moss de Oliveira and R.M. Zorzenon, {\it Physica} {\bf A231}, 504
(1996).\par

\item{[18]} P. Medawar, {\sl An Unsolved Problem of Biology}, H.K. Lewis,
London (1952).\par

\item{[19]} T.J.P. Penna, S. Moss de Oliveira and D. Stauffer, {\it Phys.
Rev.} {\bf E52}, R3309 (1995); T.J.P. Penna and S. Moss de Oliveira, {\it
J. de Physique} {\bf I5}, 1697 (1995).\par

\item{[20]} G.C. Williams, {\it Evolution} {\bf 11}, 398 (1957).\par

\item{[21]} S. Moss de Oliveira, A.T. Bernardes and J.S. S\'a Martins,
{\it Eur. Phys. J.} {\bf B7}, 501 (1999).\par

\item{[22]} T.T. Perls, E. Bubrick, C.G. Wager, J. Vijg and L. Kruglyat,
{\it The Lancet} {\bf 351}, 1560 (1998).\par

\item{[23]} P.M.C. de Oliveira, S. Moss de Oliveira, A.T. Bernardes and D.
Stauffer, {\it The Lancet} {\bf 352}, 911 (1998); --- {\it Physica} {\bf
A262}, 242 (1999).\par

\item{[24]} D. Charlesworth, M.T. Morgan and B. Charlesworth, {\it Genetic
Research} (Cambridge) {\bf 59}, 49 (1992); --- {\bf 61}, 39 (1993); ---
{\it J. Heredity} {\bf 84}, 321 (1993); F. Celada and P.Seiden, {\it
Immunology Today} {\bf 13(2)}, 56 (1992); --- {\it J. Theor. Biol.} {\bf
158}, 329 (1992); P.G. Higgs, {\it Genetic Research} (Cambridge) {\bf 63},
63 (1994); D. Alves and J.F. Fontanari, {\it Phys. Rev.} {\bf E54}, 4048
(1996); --- {\it J. Phys.} {\bf A30}, 2601 (1997); P.M.C. de Oliveira, S.
Moss de Oliveira and J.P. Radomski, {\it Theor. Biosci.} (2001); J.P.
Radomski and S. Moss de Oliveira, {\it Theor. Biosci.} (2001).\par

\bye